\documentstyle[12pt,aasms4]{article}
\def\unlock{\catcode`\@=11}
\def\lock{\catcode`\@=12}
\unlock
\def\lsim{\mathrel{\mathpalette\@versim<}}
\def\gsim{\mathrel{\mathpalette\@versim>}}
\def\@versim#1#2{\lower0.2ex\vbox{\baselineskip\z@skip\lineskip\z@skip
  \lineskiplimit\z@\ialign{$\m@th#1\hfil##\hfil$\crcr#2\crcr\sim\crcr}}}
\lock
\def\littleprime{\ifmmode{\scriptscriptstyle \prime }
    \else{\hbox{$\scriptscriptstyle \prime$ }}\fi}
\def\arcsec{\raise .9ex \hbox{\littleprime\hskip-3pt\littleprime}}
\begin{document}
 
\title{Statistics of AGN in Rich Clusters Revisited}

\author{M.J. Way\altaffilmark{1,2,3}, R.A. Flores and H. Quintana\altaffilmark{4,5}}

\affil{Department of Physics and Astronomy,
University of Missouri-St.Louis,\\
8001 Natural Bridge Rd., St.Louis, MO 63121-4499\\ I: mway@astro.princeton.edu}

\altaffiltext{1}{Visiting Astronomer, Cerro Tololo Interamerican Observatory}
\altaffiltext{2}{Visiting Astronomer, European Southern Observatory}
\altaffiltext{2}{present address: Department of Astrophysical Science, Princeton
University, Princeton, NJ 08544}
\altaffiltext{4}{Department of Astronomy and Astrophysics,
P. Universidad Catolica de Chile, Casilla 104, Santiago 22 , Chile}
\altaffiltext{5}{Presidential Chair in Science 1995}


\begin{abstract}

Using spectrophotometry of a large sample of galaxies in 19 Abell Clusters,
we have selected 42 {\it candidate} AGN by the criteria used by \cite{DTS85}
(DTS) in their analysis of the statistics of 22 AGN in 14 rich cluster fields,
which are based on the equivalent width of [OII]3727{\AA}, H$\beta$, and
[OIII]5007{\AA} emission. We have then discriminated AGN from
HII region-like galaxies (hereafter HII galaxies) in the manner developed by
\cite{VO87} (VO) using the additional information provided by H$\alpha$ and
[NII]6583{\AA} or H$\alpha$ and [SII]6716+6731{\AA} emission, in order to
test the reliability of the selection criteria used by DTS. We find that our
sample is very similar to that of DTS before we discriminate AGN from
HII galaxies, and would lead to similar conclusions. However, we find that
their method inevitably mixes HII galaxies with AGN, even for the most
luminous objects in our sample. Other authors have attempted to quantify
the relative fraction of cluster to field AGN since the study of DTS
(\cite{H93}, \cite{B97}) reaching similar conclusions, but using similar
criteria to theirs to select AGN (or using the [OIII]5007{\AA}/H$\beta$ flux
ratio test that also mixes HII galaxies with AGN).
Our sample of true AGN is left too small to reach statistically
meaningful conclusions, therefore a new study with the more time-consuming
method that includes the other lines will be required to quantify the true
relative fraction of cluster to field AGN.

\end{abstract}




\section{Introduction}
	The relative abundance of Emission Line Galaxies (ELG) in clusters
and the field has been studied by many authors as a by-product of redshift
surveys (\cite{G78}, DTS, \cite{S89}, \cite{H93}, \cite{S95},
\cite{B97}). ELG are found to be much less abundant in clusters, in
agreement with the original observation by Osterbrok (1960) about
ellipticals in the redshift survey of \cite{H56}. The early
studies concluded that this could not be due entirely to the well known
morphological segregation (\cite{D80}), but required an additional
environmental effect. However, the recent study by Biviano et al. (1997)
suggests
that this is due to a magnitude bias and that correcting for it leaves
relative abundances in agreement with morphological segregation alone,
perhaps explaining the similar finding by Moss \& Whittle (1993)
which had a bright detection threshold.

	A more difficult problem in these studies is to separate
the galaxies with Active Galactic Nuclei (AGN) or Low Ionization Nuclear
Emission Region (LINER) (both believed to be ionized by non-stellar, power-law
spectra) from those with spectra of the type of HII hot-spots in actively
star-forming regions in nearby galaxies (see e.g. \cite{O89}). A fairly
clean separation is possible if the [NII]6583{\AA}, [SII]6716+6731{\AA}
and [OI]6300{\AA} lines can be detected, as was shown by VO with
their suite of line-ratio tests based on the earlier work of
Baldwin et al. (1981).
Unfortunately, redshift surveys typically cover the wavelength range
$\sim$3500-6500{\AA} where the lines cannot be detected. Therefore, the
studies that have attempted to quantify the cluster to field ratio of
AGN/LINERs have attempted a separation based on the equivalent width $W$
of the [OIII]5007{\AA}, [OII]3727{\AA} and H$\beta$  emission lines,
following \cite{H80}. If $W_{[OIII]5007} \gsim W_{[OII]3727}$ or
$W_{H\beta} \gsim W_{[OII]3727}$, {\it and} $W_{[OIII]5007} \gsim W_{H\beta}$,
the galaxy was classified as an AGN (DTS, \cite{H93}). Biviano et al (1997)
use a criterion of [OIII] to H$\beta$ flux ratio, which was originally
thought to be a reliable indicator of AGN (\cite{SO81}). Gisler (1978) used
the Markarian lists to get AGN-type objects, but many of these are actually
HII galaxies (VO). None of these methods selects AGN/LINERs
reliably (\cite{BPT81}, VO). First, AGN and HII regions can be
cleanly separated by measuring {\it flux} ratios for different pairs of
lines. The $W_{[OIII]5007} \gsim W_{H\beta}$
criterion translates into a genuine requirement of larger [OIII]/H$\beta$
flux ratio {\it because these lines are very close to one another}
and, therefore, have the same underlying continuum. The other two criteria,
however, involve widely separated lines which may differ in $W$
{\it because of differing continua}, not differing fluxes, unless
$W >> 1$ for both lines. Second, even if the equivalent width ratios measured
flux ratios, the requirement that $[OIII]\lambda$5007 $\gsim$ H$\beta$
{\it and} $[OIII]\lambda$5007 $\gsim$ $[OII]\lambda$3727 is satisfied by
{\it both} AGN and HII regions. It is not clear that the requirement
$W_{H\beta} \gsim W_{[OII]\lambda3727}$ is not satisfied by HII regions
given the small number of HII regions in the study of \cite{H80}.
It must also be mentioned that in comparing lines such as [OII]$\lambda$3727
and [OIII]$\lambda$5007 one must take into account the slit position angle
and atmospheric differential refraction (see \cite{F82}) which
can place more flux in one line versus another simply due to the position
of the slit on the sky. This effect is stronger when lines are widely
separated {\it unlike} H$\beta$ and [OIII]$\lambda$5007.
 
	We have attempted, however, to calibrate the technique of equivalent
widths by selecting {it candidate} AGN from a large sample of galaxies in 19
Abell Clusters for which spectra were obtained as part of a redshift survey.
We have used the criteria used by DTS to identify 42 {it candidate} objects.
Using 3 nights in 1993 on the ESO 2.2 meter and 4 nights in 1995 at
the CTIO 1.5m we obtained flux calibrated spectra covering the wavelength
range 4200--7500{\AA} for our 42 {it candidate} AGN. By measuring accurate
ratios of [OIII]$\lambda$5007/H$\beta$, [SII]$\lambda$6716+6731/H$\alpha$ and
[NII]$\lambda$6583/H$\alpha$ we have then separated AGN/LINERs from
HII galaxies using the criteria developed by VO. We find that no
threshold in equivalent width can cleanly separate them. We do not find it
possible to separate them by an absolute luminosity threshold either,
contrary to what DTS argued: our AGN/LINERs and HII galaxies are mixed at
all luminosities, in agreement with studies of nearby ELG that also find
quite luminous HII galaxies (\cite{H96}, \cite{G97}).
Thus, it is fair to say that the relative fraction
of cluster to field AGN is not presently known, as our study is left with
too small a sample to reach statistically sound conclusions. A study with
the more time-consuming method that includes the other lines will be required
to adress this question, but a step in the right direction is the recent
study of {\it field} AGN based on the CFRS (\cite{T96}).

	We present the observations and data reduction for the 42
{it candidate} AGN in the next section, and then the analysis of the reduced
data in the subsequent section. Finally, we close with a section
of discussion of the analysis, and the conclusions we draw from it.
We use a Hubble constant H$_{o}$ = 100 h km/s/Mpc.


\section{Observations and Reductions}
	We started by selecting {\it candidate} AGN from a large set of ELG
spectra obtained as part of a redshift survey in the field of 19 Abell
Clusters (see \cite{QIW97}, \cite{WQI97}, \cite{QRW96}), using the criteria
of DTS for selection. Spectra
for these objects, covering the wavelength range $\sim$4200--7500{\AA},
were then obtained from two different observing runs.

	The first run was at the ESO-MPI 2.2 meter telescope using EFOSC2 on
the 2 nights of October 18 and 19, 1993. The set up was with
CCD \#19, a 1024X1024 Thompson chip with 19$\micron$ pixels which has
a gain of 2e$^{-}$/ADU with readout noise of $<$5e$^{-}$. Grism \#6 with
300 lines/mm blazed at 5000{\AA} and a dispersion of $\sim$2.7{\AA}/pixel,
and a 1.5$\arcsec$ slit.  This yielded a resolution of $\sim$9{\AA} FWHM
and a wavelenth coverage of $\sim$4600-7200{\AA}.

	The second run was on
the CTIO 1.5 meter telescope using the the Cassegrain focus Boller
and Chivens spectrograph on the 4 nights of November 29 to December 2, 1995.
A Loral 1200X800 CCD was used which has 15$\micron$ pixels. Gain was set to
1e$^{-}$/ADU which as a read noise of 5.88e$^{-}$.  Grating \#32 was used
which has 300 lines/mm blazed at 6750{\AA} giving a wavelength
coverage of $\sim$3450{\AA}.  This in turn gave us excellent sensitivity 
in the 4200-7500{\AA} range used and a dispersion of $\sim$2.8{\AA}/pixel.
4.5$\arcsec$ = 221.2$\micron$ and 4.0$\arcsec$ = 221.2$\micron$ slits were
used for object exposures. These gave resolutions of $\sim$8-9{\AA} FWHM.
Helium, Neon, and Argon lamps were also used to generate comparison frames for
wavelength calibrating the data. See Table 1 for a summary of telescope
and instrument properties.

{\scriptsize
\begin{deluxetable}{lcccccccc}
\tablenum{1}
\tablecaption{Spectra Observed}
\tablehead{
\colhead{Telescope}              & \colhead{CCD}                & 
\colhead{e$^{-}$/ADU}            & \colhead{Read Noise e$^{-}$} & 
\colhead{Blaze({\AA})}           & \colhead{l/mm}               &
\colhead{{\AA}/pixel}            & \colhead{Res({\AA})}  &
\colhead{Range({\AA})} }
\startdata 
ESO 2.2m &1024$^{2}$ Thompson & 2 & 5    & 5000 & 300 & 2.7 & 9   & 4600-7200\nl
CTIO 1.5m&1200X800 Loral      & 1 & 5.88 & 6750 & 300 & 2.8 & 8-9 & 4200-7500\nl
\enddata
\end{deluxetable} 
}

	Six spectra were obtained from the ESO 2.2 meter run and
another forty from the CTIO 1.5 meter. Some objects were observed more than
once and, when appropriate, the spectrum of better S/N was selected for the
analysis described in the next section. The spectra were calibrated
with the spectrophotometric standards of Baldwin \& Stone (1984) and
Hamuy et al. (1992).
Spectrophotometric standards were taken several times over the course of
the night to assure good airmass coverage. Three equal length
exposures were taken of each object. Each set of 3 exposures were later
combined with a median filter to eliminate cosmic rays using the IRAF
COMBINE task.

	The reduction of each data run was basically the same. Since we used
the IRAF (\cite{T93}) software package to reduce our data we had to update
the EFOSC2 fits headers with some additional information required by IRAF.
Bias images were summed with the ZEROCOMBINE task, flat-fields
with FLATCOMBINE, and each of the three object copies with the COMBINE task
using a median filter to rid them of cosmic rays.
Using the RESPONSE task the combined flat-field was then fit with a 6th
order spline to fit inherent large scale features.
CCDPROC was then used to bias and flatfield correct the object spectra
and spectrophotometric standards. Afterwards CTIOSLIT was used to extract
the object and spectrophotometric standard spectra from the ccd frames.
It then uses the spectrophotometric standards to correct the object spectra
for extinction and flux. Not all of our data was successfully flux
calibrated (due to some non-photometric weather), but for the
purposes of this paper they are more than adequate. This is because
the emission lines we wish to measure are ratioed with respect to close
neighboring lines, and therefore the shape of the underlying continuum
is not crucial, e.g. H$\beta$ and [OIII]5007{\AA} or H$\alpha$ and
[NII]6583{\AA}.

	An automated task in IRAF called FITPROFS was used to
measure the flux within the desired line profiles. This task allows
one to set the region about the line from which it can pick
the continuum. The continuum is then fit with a linear function
in the region under the line we want to measure.
Once the region is chosen, one also inputs the center of the line
to be measured, and a gaussian was chose to fit the line profile with. To
estimate the error in the emission line fit one must estimate the
uncertainty of a pixel in the line profile (which is what FITPROFS requires
in one of it's parameters).  This can be obtained by estimating the
RMS on a piece of the spectrum free of emission or absorbtion lines (or else
a sigma clipping of any emission or absorption lines).


\section{Analysis}
	The basic data for our {it candidate} AGN, before classifying them by
the diagnostic diagrams constructed from the data described in section 2,
are presented in Table 2. They were selected from the ELG spectra of a
redshift survey in the field of 19 Abell Clusters that are representative
of clusters of richness $R \ge 1$ in the Abell catalog, and therefore
similar to the cluster sample of Dressler \& Shechtman (1988) used by DTS.
The sole exception
is A133, which is a $R = 0$ cluster, but whose velocity dispersion is more
like that of a $R = 1$ cluster (\cite{WQI97}).

	The range of absolute magnitudes of our {it candidate} AGN is similar
to that of the DTS sample, and so is the distribution of magnitudes. The main
difference is that our objects are distributed out to much larger distances.
Given the similarities of the two samples, it seems fair to conclude that
we have been successful at obtaining a sample very similar to that of the
DTS study. It is not surprising, then, that the fraction of candidates
in and out of clusters are rather similar: here 64\% and 36\% respectively,
compared to 59\% and 41\% in the DTS study. If A133 were not included on
account of being a $R = 0$ cluster, then our sample becomes even more like
the DTS sample, with the percentages above changing to 60\% and 40\% for
our sample. Thus, {\it we would have been lead to the same conclusions with
regards to the relative fraction of cluster to field AGN of the many studies
that have identified AGN in a manner that does not fully separate them from
HII galaxies}.

	We can now turn to the analysis of our longer wavelength data to
separate AGN/LINERs from HII galaxies. The basic data obtained from the
reduction described in section 2 are presented in Table 3, and the tabulated
flux ratios are plotted in the diagnostic diagrams of VO in Figure
1. The dashed line in each diagram is the line that VO found
to provide a good empirical separation of nearby AGN/LINERs and HII regions,
and we shall use it here to distinguish AGN/LINERs from HII galaxies in our
sample. Tresse et al. (1997) use an extreme theoretical model to obtain
an upper
limit to the HII boundary in the diagnostic diagrams which differs
significantly from the VO empirical boundaries for [OIII]/H$\beta$
$\lsim 3$. However, for the diagnostic diagram that can be compared directly
(Fig. 1(b)), we find that this would affect only a few of the objects we have
classified as transitional or uncertain anyway.
Our flux ratios are not corrected for reddening, but such a correction is
small for the lines chosen in the diagnostic diagrams (VO). We have
not corrected for stellar absorption under the Balmer lines either, although
this is a more substantial correction. Since this only moves the objects down
and a bit to the left in the diagrams, it can only decrease the number of
AGN/LINERs in our sample and, therefore, not affect our main conclusion that
blue-spectra based studies of AGN using equivalent width selection criteria
(or [OIII]/H$\beta$ flux ratio) get a large ``contamination'' from HII
galaxies. There are sizeable errors associated with some of our flux ratios.
This is because the lines are fairly weak in some of the objects, and some of
the lines are weak in nearly all of the objects.
For clarity, we show separately in Figure 2
the same diagnostic diagrams of Fig. 1 with our estimate of $1\sigma$ errors.
Note that despite the substantial errors, most of the points stay entirely
on the same side of the empirical dividing line. Therefore, we conclude that
a large number of our objects do have spectra characteristic of HII galaxies.

On the basis of the diagnostic diagrams, Fig. 1, we have classified the
galaxies in our sample in two ways. First, a subjective classification based
on inspection of the the distances of the flux-ratio data points to the
empirical dividing lines, which we list in Table 3 for each of the diagnostic
diagrams. The dividing lines can be moved, we estimate, by $\pm 0.1$ without
affecting the classifications of VO, therefore we consider the
distance to be significant only if it exceeds $0.1$ in absolute value. The
identification is considered secure if all distances are of the same sign
(unless they are all $<0.1$), or if two distances are of the same sign, but
the third is $<0.1$ in absolute value.
Objects for which two distances are large and of opposite sign make an object
``transitional'' or ``uncertain'', unless the third
distance is very much larger. Not every object is consistently in the same
region of each of the diagrams in Fig. 1, therefore we list the distance in
Table 3 with an appropriate sign to indicate if the galaxy lies in the
AGN/LINER ($+$) or HII galaxy ($-$) region in each of the diagrams. For the
objects
classified as AGN/LINER or HII galaxy we have inspected the individual spectra
to make sure that they are consistent with the classification of the object.
Second, we have also classified the objects by a simple, objective rule: if
the sum of the distances, columns 6-8 of Table 3, exceeds $0.3$, then the
object
is classified as an AGN/LINER or HII galaxy depending on the sign, and
objects in between are classified as transitional. We show in Figure 3 the
sum of the distances, column 9 of Table 3, plotted against recessional
velocity, column 8 of Table 2. Filled squares are AGN/LINERS selected
subjectively as explained above, open squares are transitional objects, and
stars are HII-region objects. One can see that if we had classified galaxies
with the objective criterion (dotted line) the results would have been
essentially the same. Also, it is apparent in the figure that we do not have
systematic effects related to distance affecting our classifications.

Our final classification, listed in column 10 of Table 3, is based on the
objective criterion, but with the $\pm 0.1$ uncertainty treated as statistical.
Thus, if the sum of the distances, column 9 of Table 3, is $\Sigma d_i >
\sqrt{3} \times 0.1$ ($\Sigma d_i < -\sqrt{3} \times 0.1$,
$-\sqrt{3} \times 0.1 < \Sigma d_i < \sqrt{3} \times 0.1$), we have
classified the object as an AGN/LINER (HII galaxy, transitional) and
denoted them as S/L (HII, T) in Table 3. This does not change at all the
number of objects that are classified as AGN/LINERs. Two objects (C3141-34
and C3864-5b) had to be classified differently because of undectectable flux
in some lines. C3141-34 had no dectectable [NII] and [SII] lines, but it was
detected in H$\alpha$. Therefore it clearly belongs in the HII category
given its [OIII]/H$\beta$ flux ratio, and this is confirmed by the observed
[OI]/H$\alpha$ ratio. C3864-5b was not detected in H$\beta$, but for any
[OIII]/H$\beta$ flux ratio the other flux ratios put it in the AGN/LINER
region. All but one transitional
objects have two distances of opposite sign making a small $\Sigma d_i$. It
may be that these are objects in which there is star formation surrounding
a LINER (see \cite{H96}; the exact definition of ``transitional'' is not the
same), which would be picked up because our slits cover a few kiloparsecs
around the galaxy center. However, three of them have rather uncertain
flux ratios and, therefore, it may be that the small $\Sigma d_i$ is due
to noise.

With the classification assigned in Table 3, we have looked for equivalent
width thresholds that might help distinguish AGN/LINERs from HII regions.
However, no threshold in the equivalent width ratios $W_{[OIII]}/W_{[OII]}$
or $W_{H\beta}/W_{[OII]}$, and $W_{[OIII]}/W_{H\beta}$ can separate them.
Likewise, we have tried to see if absolute luminosity can help distinguish
them. Unfortunately, AGN/LINERs and HII galaxies in our sample remain
mixed at all luminosities, even though the ratio of HII galaxies to
AGN/LINERs indeed decreases with luminosity.


\section{Discussion and Conclusions}
	The analysis of the previous section leaves us, unfortunately, with
only 10 AGN/LINERs. Naturally, since our slits cover a few kiloparsecs
around the galaxy centers, we cannot be sure that the emission from our
AGN/LINERs is really from their nuclei. However, we are not aware of any
detection of their type of spectra from non-nuclear regions in studies of
nearby ELG. Our main conclusion is that blue-spectra based studies of AGN
using equivalent width selection criteria (or [OIII]/H$\beta$ flux ratio)
get a large ``contamination'' from HII galaxies, and therefore that the
true relative fraction of cluster to field AGN is not known.

	To the extent that both AGN and HII galaxies depend on the
availability of gas for their existence, one might expect this ratio to
be simply that determined from studies of ELG that do not separate between
these classes. However, different spectral types are found to be
distributed differently as a function of radial distance in a detailed study
of one cluster (\cite{F97}). Another reason it would be
interesting to separately study HII galaxies from AGN in clusters is that
it would be possible to clarify if indeed some clusters have an abnormally
high incidence of AGN (DTS), and thereby whether environmental effects are
at work. Two of the clusters (A133 and A3194) in our sample show several
{\it candidate} AGN ($\sim 5\%$ of the cluster galaxies), as was the case of
one cluster in the DTS sample. However, in the case of A133 the flux ratio
tests indicate that only one is (marginally) an AGN/LINER. In the case of
A3194 all three {\it candidate} AGN turn out to be bona fide cluster
AGN/LINERs, although one of them (C3194-7c) is rather far from the cluster
center (at least $2.2 h^{-1}$ Mpc). We cannot know, of course, if this is
a high fraction of AGN for clusters without knowing the true abundance of
AGN in clusters.

\acknowledgements 
This work has been supported by an NSF grant and by Research and Research
Board awards at The University of Missouri--St.Louis. One of us (RF)
would like to acknowledge the hospitality of the Physics Department at
UCSC while this work was being written, and D. Osterbrok for a useful
discussion on AGN. This research has made use of NASA's Extragalactic
Database, ADS Abstract Service, and the Digitized Sky Survey at STScI.
This project was partially supported by FONDECYT grants 8970009 and 7960004.


{\scriptsize
\begin{deluxetable}{lrlclllrccc}
\tablenum{2}
\tablecaption{Objects Observed}
\tablehead{
\colhead{ID\tablenotemark{(1)}} &
\colhead{$\alpha$ (1950)\tablenotemark{(2)}} & 
\colhead{$\delta$ (1950)\tablenotemark{(2)}} &
\colhead{mag\tablenotemark{(3)}} &
\colhead{$W_{[OII]}$\tablenotemark{(4)}} &
\colhead{$W_{H\beta}$\tablenotemark{(4)}} &
\colhead{$W_{[OIII]}$\tablenotemark{(4)}} &
\colhead{v$_{\odot}$\tablenotemark{(5)}} &
\colhead{Mag\tablenotemark{(6)}} &
\colhead{R\tablenotemark{(7)}} & \colhead{Memb\tablenotemark{(8)}}}
\startdata 
C2854-74  &  1:00:15 & -51:17:57.1  & 16.2 & 5.273 & 9.942  & 8.306 & 21074 & -21.9 & 1 & N \nl
C487-51   &  4:21:34 & -24:13:32.3  & 15.1 & 34.49 & 7.938  & 85.94 & 17726 & -22.6 & 1 & N \nl
C3194-28  &  3:57:20 & -30:20:38.3  & 16.1 & 33.91 & 1.278  & 52.4  & 29278 & -22.7 & 2 & Y \nl
C3194-7c  &  3:55:13 & -30:26:53.6  & 16.6 & 38.73 & 2.229  & 8.386 & 28692 & -22.2 & 2 & Y \nl
C3194-31  &  3:57:09 & -30:26:39.8  & 16.4 & 146.3 & 7.511  & 50.26 & 27895 & -22.3 & 2 & Y \nl
C3264-4d  &  4:27:28 & -49:16:53.5  & 16.3 & 75.51 & 11.85  & 42.4  & 29047 & -22.5 & 1 & Y \nl
C3264-5e  &  4:27:16 & -49:18:04.4  & 17.7 & 49.17 & 39.71  & 91.31 & 31900 & -21.3 & 1 & N \nl
C2854-45  &  0:58:07 & -50:32:59.2  & 17.2 & 100.9 & 96.90  & 110   & 18719 & -20.7 & 1 & Y \nl
C2854-68  &  0:58:06 & -50:55:33.3  & 17.1 & 34.35 & 12.07  & 20.43 & 19144 & -20.8 & 1 & Y \nl
C3153-67  &  3:36:51 & -34:38:05.4  & 16.7 & 8.411 & 18.57  & 28.8  & 37184 & -22.7 & 1 & Y \nl
C3153-29  &  3:41:19 & -34:42:58.3  & 16.4 & 8.036 & 4.425  & 10.99 & 37536 & -23.0 & 1 & Y \nl
C3223-46  &  4:01:56 & -30:13:36.8  & 17.1 & 22.86 & 3.88   & 13.82 & 18246 & -20.7 & 2 & Y \nl
C3223-6a  &  4:01:14 & -30:35:49.5  & 17.0 & 6.763 & 3.112  & 2.925 & 18193 & -20.8 & 2 & Y \nl
C3223-14  &  4:03:30 & -30:26:13.3  & 17.4 & 25.62 & 3.203  & 10.59 & 17573 & -20.3 & 2 & Y \nl
C3223-01  &  4:02:43 & -30:01:31.8  & 17.3 & 15.55 & 14.0   & 41.69 & 17909 & -20.5 & 2 & Y \nl
C487-38   &  4:22:06 & -24:51:48.4  & 17.6 & 26.05 & 4.322  & 12.29 & 17268 & -20.1 & 1 & N \nl
C2871-6c  &  1:04:58 & -37:02:12.4  & 15.8 & 24.54 & 3.584  & 27.44 &  3932 & -18.7 & 2 & N \nl
C3921-0d  & 22:48:51 & -64:09:56.0  & 17.2 & 36.58 & 13.28  & 25.59 & 31999 & -21.8 & 2 & N \nl
C2871-53  &  1:05:19 & -36:59:02.0  & 17.1 & 13.08 & 13.30  & 10.06 & 35870 & -22.2 & 2 & Y \nl
C2923-71  &  1:30:48 & -31:51:41.8  & 17.6 & 39.34 & 9.348  & 28.87 & 25540 & -20.9 & 1 & N \nl
C3142-53  &  3:34:49 & -39:43:22.4  & 17.4 & 9.494 & 4.718  & 2.40  & 21509 & -20.8 & 1 & N \nl
C3142-08  &  3:35:13 & -39:39:43.3  & 17.5 & 40.97 & 10.83  & 30.62 & 17530 & -20.2 & 1 & N \nl
C3223-26  &  4:04:05 & -30:32:32.0  & 15.9 & 5.876 & 2.495  & 1.225 & 17613 & -21.8 & 2 & Y \nl
C3188-62  &  3:55:40 & -27:17:52.0  & 17.4 & 24.52 & 10.02  & 24.99 & 21447 & -20.8 & 1 & N \nl
C3188-16  &  3:57:07 & -27:12:16.8  & 17.3 & 33.83 & 4.079  & 17.67 & 18772 & -20.6 & 1 & Y \nl
C3141-33  &  3:34:01 & -28:54:45.0  & 17.6 & 8.839 & 0.7968 & 13.39 & 30721 & -21.3 & 1 & Y \nl
C3864-5b  & 22:12:52 & -52:45:32.0  & 17.1 & 9.28  & 0.0    & 10.56 & 15936 & -20.4 & 1 & N \nl
C3112-0b  &  3:16:43 & -44:30:21.3  &      & 64.21 & 18.40  & 80.49 & 22582 &       & 2 & Y \nl
C2911-75  &  1:24:02 & -38:27:53.6  & 17.5 & 41.08 & 6.678  & 17.26 & 24201 & -20.9 & 1 & Y \nl
C3112-25  &  3:16:27 & -44:49:55.3  &      & 10.20 & 5.694  & 2.768 & 20085 &       & 2 & N \nl
C3141-04  &  3:35:07 & -27:36:04.6  & 17.8 & 24.30 & 3.599  & 11.29 & 31338 & -21.2 & 1 & Y \nl
C3141-34  &  3:34:29 & -28:50:18.3  & 17.8 &  0.0  & 25.76  & 8.082 & 38857 & -21.7 & 1 & N \nl
C3151-06  &  3:40:54 & -28:38:12.7  & 17.0 & 55.46 & 7.753  & 17.06 & 20777 & -21.1 & 1 & Y \nl
C3151-14  &  3:38:48 & -28:51:54.8  & 16.8 & 47.65 & 12.26  & 23.01 & 17778 & -21.0 & 1 & N \nl
C3266-12  &  4:32:52 & -61:04:44.5  &      & 11.30 & 2.10   & 6.30  & 18193 &       & 2 & Y \nl
C3223-78  &  4:05:30 & -31:42:19.35 & 17.3 & 58.98 & 13.45  & 31.02 & 21104 & -20.8 & 2 & N \nl
E3266-03  &  4:33:58 & -61:00:14.8  &      & 7.80  & 3.27   & 13.59 & 17671 &       & 2 & Y \nl
E133-29   &  0:59:40 & -22:22:25.3  & 16.1 & 12.55 & 2.70   & 26.35 & 16403 & -21.5 & 0 & Y \nl
E133-30   &  0:59:46 & -22:08:16.2  & 15.9 & 12.20 & 1.82   & 15.10 & 16315 & -21.7 & 0 & Y \nl
E133-12   &  1:01:40 & -21:53:04.3  & 16.5 & 9.00  & 4.40   & 3.66  & 16228 & -21.0 & 0 & Y \nl
E133-33   &  0:59:41 & -22:20:49.3  & 18.1 & 6.93  & 3.10   & 2.10  & 15619 & -19.4 & 0 & Y \nl
E119-08   &  0:56:14 & -01:21:59.7  & 15.7R& 18.48 & 4.07   & 41.57 & 13935 & -21.5R& 1 & Y \nl
\tablenotetext{(1)}{Object identification: prefix (E = ESO 1993 run,
C=CTIO 1995 run) followed by Abell cluster number and a fiber id
number from redshift survey.}
\tablenotetext{(2)}{Right ascension ($\alpha$) and declination ($\delta$).}
\tablenotetext{(3)}{Apparent magnitude in V, except for one available
only in R; some are not available.}
\tablenotetext{(4)}{Equivalent width, in \AA, for [OII] (H$\beta$, [OIII])
line (from redshift survey fiber spectra).}
\tablenotetext{(5)}{Heliocentric recession velocity, in km/s.}
\tablenotetext{(6)}{Absolute magnitude in V for h = 0.5 as in DTS.}
\tablenotetext{(7)}{Cluster richness class.}
\tablenotetext{(8)}{Membership in cluster based on velocity distribution,
as in DTS.}
\enddata
\end{deluxetable} 
}

{\scriptsize
\begin{deluxetable}{lcccccccccc}
\tablenum{3}
\tablecaption{Diagnostic Flux Ratios}
\tablehead{
\colhead{ID\tablenotemark{(1)}} &
\colhead{$\frac{{\rm [NII]}}{{\rm H}\alpha}$\tablenotemark{(2)}} &
\colhead{$\frac{{\rm [SII]}}{{\rm H}\alpha}$\tablenotemark{(2)}} &
\colhead{$\frac{{\rm [OIII]}}{{\rm H}\beta}$\tablenotemark{(2)}} &
\colhead{$\frac{{\rm [OI]}}{{\rm H}\alpha}$\tablenotemark{(2)}} &
\colhead{$d_{\rm [NII]}$\tablenotemark{(3)}} &
\colhead{$d_{\rm [SII]}$\tablenotemark{(3)}} &
\colhead{$d_{\rm [OI]}$\tablenotemark{(3)}} &
\colhead{$\Sigma d_i$\tablenotemark{(4)}} &
\colhead{Class\tablenotemark{(5)}} &
\colhead{r (\arcmin)\tablenotemark{(6)}}}
\startdata
C2854-74 & -0.1979 & -1.1465 &  0.0244 & -1.6831 & 0.0509 & -.7113 &  -.4572 & -1.1176 & HII & 39 \nl
C487-51  &  0.0064 & -0.2926 &  1.0640 & -1.1020 & 0.6285 & 0.4971 &  0.4503 &  1.5759 & S/L & 16 \nl
C3194-28 &  0.0765 & -0.1329 &  1.3320 & -0.8865 & 0.8557 & 0.7922 &  0.7740 &  2.4219 & S/L &  3 \nl
C3194-7c &  0.1200 &  0.1027 &  0.6504 & -0.9782 & 0.5167 & 0.6581 &  0.3944 &  1.5692 & S/L & 31 \nl
C3194-31 & -0.0068 & -0.0358 &  0.6523 & -0.8139 & 0.4010 & 0.5285 &  0.5501 &  1.4796 & S/L &  8 \nl
C3264-4d & -0.7483 & -0.3626 &  0.3858 & -1.4202 & -.3810 & 0.1332 &  -.1086 & -0.3564 & HII & 40 \nl
C3264-5e & -1.1287 & -0.7455 &  0.6949 & -1.0142 & -.3753 & -.0870 &  0.3757 & -0.0870 &  T  & 43 \nl
C2854-45 & -0.8767 & -0.4712 &  0.3053 & -1.1397 & -.5326 & 0.0065 &  0.1363 & -0.3898 & HII & 17 \nl
C2854-68 & -0.8104 & -0.4199 &  0.1774 & -1.4684 & -.5216 & 0.0299 &  -.2135 & -0.7052 & HII & 11 \nl
C3153-67 & -0.1349 & -0.7997 &  0.4898 & -1.3594 & 0.2204 & -.2405 &  -.0180 & -0.0381 &  T  & 36 \nl
C3153-29 & -0.2001 & -0.7859 &  0.3038 & -1.9484 & 0.0974 & -.2949 &  -.6351 & -0.8326 & HII & 38 \nl
C3223-46 & -0.6998 & -0.3880 &  0.3638 & -1.3326 & -.3491 & 0.1025 &  -.0319 & -0.2785 & HII & 82 \nl
C3223-6a & -0.2537 & -0.2649 &  0.4659 & -1.2803 & 0.1011 & 0.2506 &  0.0487 &  0.4004 & S/L & 83 \nl
C3223-14 & -1.0770 & -0.4298 &  0.4739 & -0.8013 & -.5471 & 0.0973 &  0.5078 &  0.0580 &  T  & 56 \nl
C3223-01 & -0.2196 & -0.6494 &  0.5407 & -1.5666 & 0.1612 & -.0826 &  -.1946 & -0.1160 &  T  & 80 \nl
C487-38  & -0.8258 & -0.2322 &  0.2806 & -1.2759 & -.4984 & 0.2332 &  -.0009 & -0.2661 & HII & 37 \nl
C2871-6c & -0.8397 & -0.4264 &  0.3076 & -1.2380 & -.4993 & 0.0503 &  0.0422 & -0.4068 & HII & 10 \nl
C3921-0d & -0.8324 & -0.4774 &  0.0637 & -1.4240 & -.5741 & -.0425 &  -.1956 & -0.8122 & HII & 46 \nl
C2871-53 & -0.4386 & -0.9343 & -0.1271 & -1.3006 & -.2013 & -.5156 &  -.1077 & -0.8246 & HII &  4 \nl
C2923-71 & -0.6897 & -0.5551 &  0.3586 & -0.4273 & -.3422 & -.0585 &  0.8404 &  0.4397 & S/L & 33 \nl
C3142-53 & -0.4091 & -0.5468 & -0.1061 & -1.1723 & -.1705 & -.1277 &  0.0218 & -0.2764 & HII & 15 \nl
C3142-08 & -0.5680 & -0.5819 &  0.5168 & -1.4641 & -.1585 & -.0299 &  -.1073 & -0.2957 & HII & 19 \nl
C3223-26 & -0.3155 & -0.6981 & -0.0608 & -1.3718 & -.0737 & -.2746 &  -.1674 & -0.5157 & HII & 45 \nl
C3188-62 & -0.6692 & -0.3550 &  0.7318 & -1.0153 & -.0978 & 0.2653 &  0.3876 &  0.5551 & S/L &  7 \nl
C3188-16 & -0.7074 & -0.3883 &  0.3263 & -1.4281 & -.3716 & 0.0920 &  -.1341 & -0.4137 & HII & 21 \nl
C3141-33 & -0.0944 & -0.3770 &  1.1109 & -1.5080 & 0.5809 & 0.4629 &  0.1353 &  1.1791 & S/L & 44 \nl
C3864-5b &  0.2815 & -0.1282 &         & -0.0893 &        &        &         &$>$1.75  & S/L & 62 \nl
C3112-0b & -1.0281 & -0.7460 &  0.6580 & -1.9838 & -.3617 & -.1094 &  -.5126 & -0.9837 & HII &  9 \nl
C2911-75 & -0.9670 & -0.5264 &  0.3438 & -1.4773 & -.5809 & -.0357 &  -.1759 & -0.7925 & HII & 14 \nl
C3112-25 & -0.3402 & -0.5732 & -0.3764 & -1.7861 & -.1173 & -.1725 &  -.6263 & -0.9161 & HII & 25 \nl
C3141-04 & -0.6283 & -0.4655 &  0.0600 & -1.0901 & -.3728 & -.0311 &  0.1309 & -0.2730 & HII & 37 \nl
C3141-34 &         &         & -0.4162 & -1.8316 &        &        &  -.6769 &         & HII & 38 \nl
C3151-06 & -0.9735 & -0.3411 &  0.7438 & -0.8557 & -.2603 & 0.2831 &  0.5414 &  0.5642 & S/L & 40 \nl
C3151-14 & -0.8660 & -0.5877 &  0.4450 & -1.2858 & -.4382 & -.0611 &  0.0369 & -0.4624 & HII &  6 \nl
C3266-12 & -0.2505 & -0.4134 &  0.2764 & -1.3456 & 0.0413 & 0.0554 &  -.0693 &  0.0274 &  T  & 47 \nl
C3223-78 & -0.8164 & -0.5135 &  0.3096 & -1.3439 & -.4776 & -.0331 &  -.0586 & -0.5693 & HII & 48 \nl
E3266-03 & -0.1366 & -0.2371 &  1.0840 & -0.6779 & 0.5315 & 0.5532 &  0.8358 &  1.9205 & S/L & 63 \nl
E133-29  & -0.2388 & -0.6129 &  0.9500 & -2.0615 & 0.3665 & 0.1806 &$>$-.4284&  0.1187 &  T  & 20 \nl
E133-30  & -0.1661 & -0.3583 &  0.9886 & -2.1787 & 0.4470 & 0.3989 &$>$-.5334&  0.3125 & S/L &  8 \nl
E133-12  & -0.4133 & -0.5030 & -0.2431 & -2.1696 & -.1833 & -.0940 &  -.9842 & -1.2615 & HII & 25 \nl
E133-33  & -0.3738 & -0.4022 & -0.2557 & -1.7467 & -.1446 & 0.0059 &  -.5686 & -0.7073 & HII & 19 \nl
E119-08  &  0.1194 &  0.1324 &  1.2810 & -0.7206 & 0.8539 & 0.9675 &  0.8891 &  2.7105 & S/L & 38 \nl
\tablenotetext{(1)}{Object identification, as in Table 2.}
\tablenotetext{(2)}{Logarithm of [NII]/H$\alpha$
([SII]/H$\alpha$, [OIII]/H$\beta$, [OI]/H$\alpha$) flux ratio.}
\tablenotetext{(3)}{Distance from dividing line in the [OIII]/H$\beta$
vs. [NII]/H$\alpha$ ([SII]/H$\alpha$, [OI]/H$\alpha$) diagnostic
diagram, Fig. 1a (1b, 1c).}
\tablenotetext{(4)}{Sum of distances in columns 6, 7 and 8.}
\tablenotetext{(5)}{Final classification; see text.}
\tablenotetext{(6)}{Projected distance from cluster center, in minutes
of arc.}
\enddata
\end{deluxetable}
}
\clearpage
\begin{figure}

\caption{Diagnostic diagrams for our {\it candidate} AGN, showing the logarithm
of the [OIII]/H$\beta$ flux ratio vs (a) [NII]/H$\alpha$, (b) [SII]/H$\alpha$
and (c) [OI]/H$\alpha$ flux ratios. The dashed line in each diagram is the
empirical dividing line separating AGN/LINERs from HII-regions and
HII-region-like galaxies in the study of VO.}

\caption{Same as Figure 1, with $1\sigma$ error bars added.}

\caption{Comparison of the two classifications discussed in the text.
The quantity $\Sigma d_i$ (column 9 of Table 3) is plotted against the
recession velocity (column 8 of Table 2) for each galaxy. The filled squares
(open squares, stars) are those classified as AGN/LINERs (transitional,
HII galaxy) based on the subjective criterion discussed. The dotted line
marks a possible separation based simply on $\Sigma d_i$ (see text).}

\end{figure}
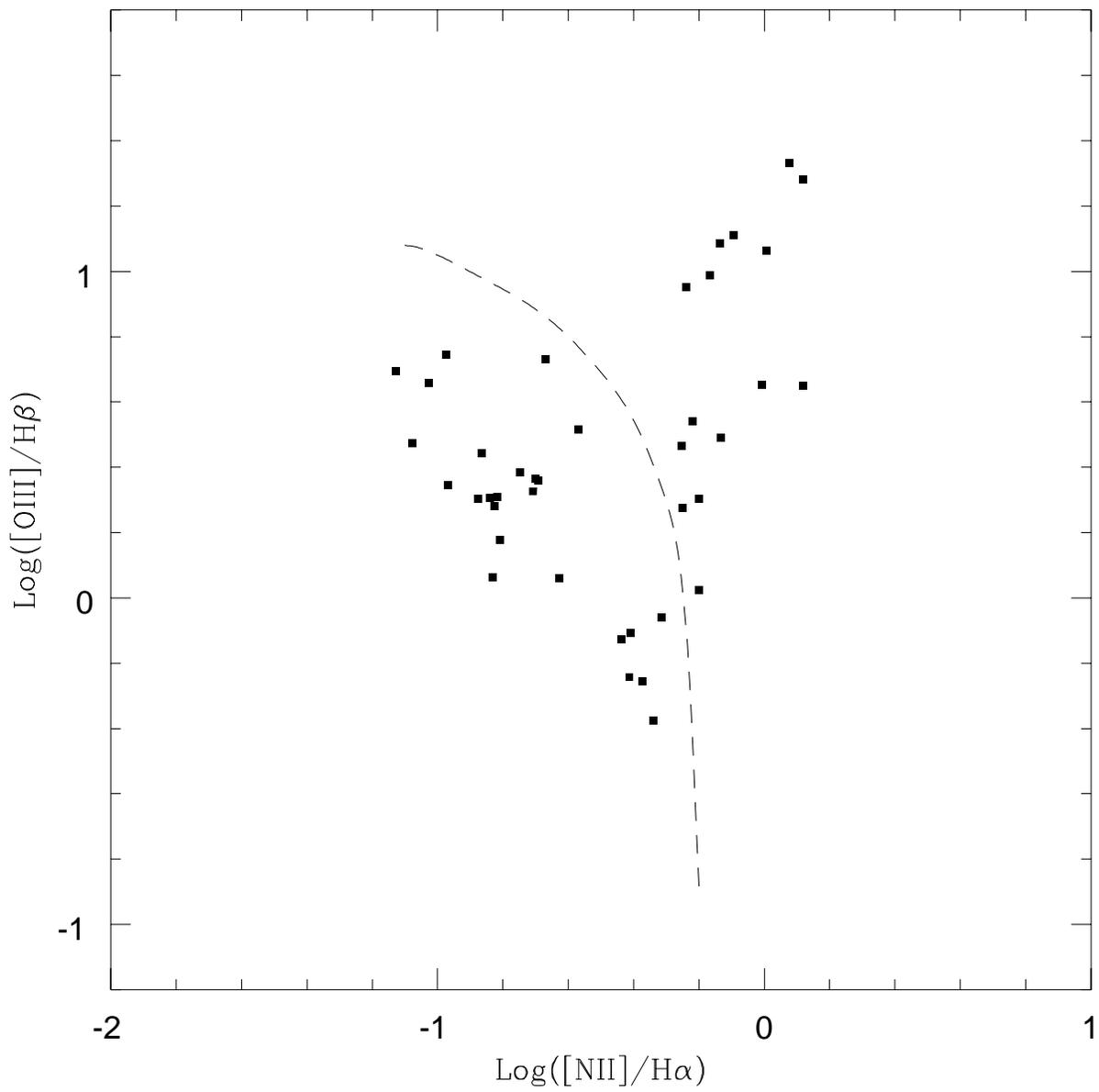
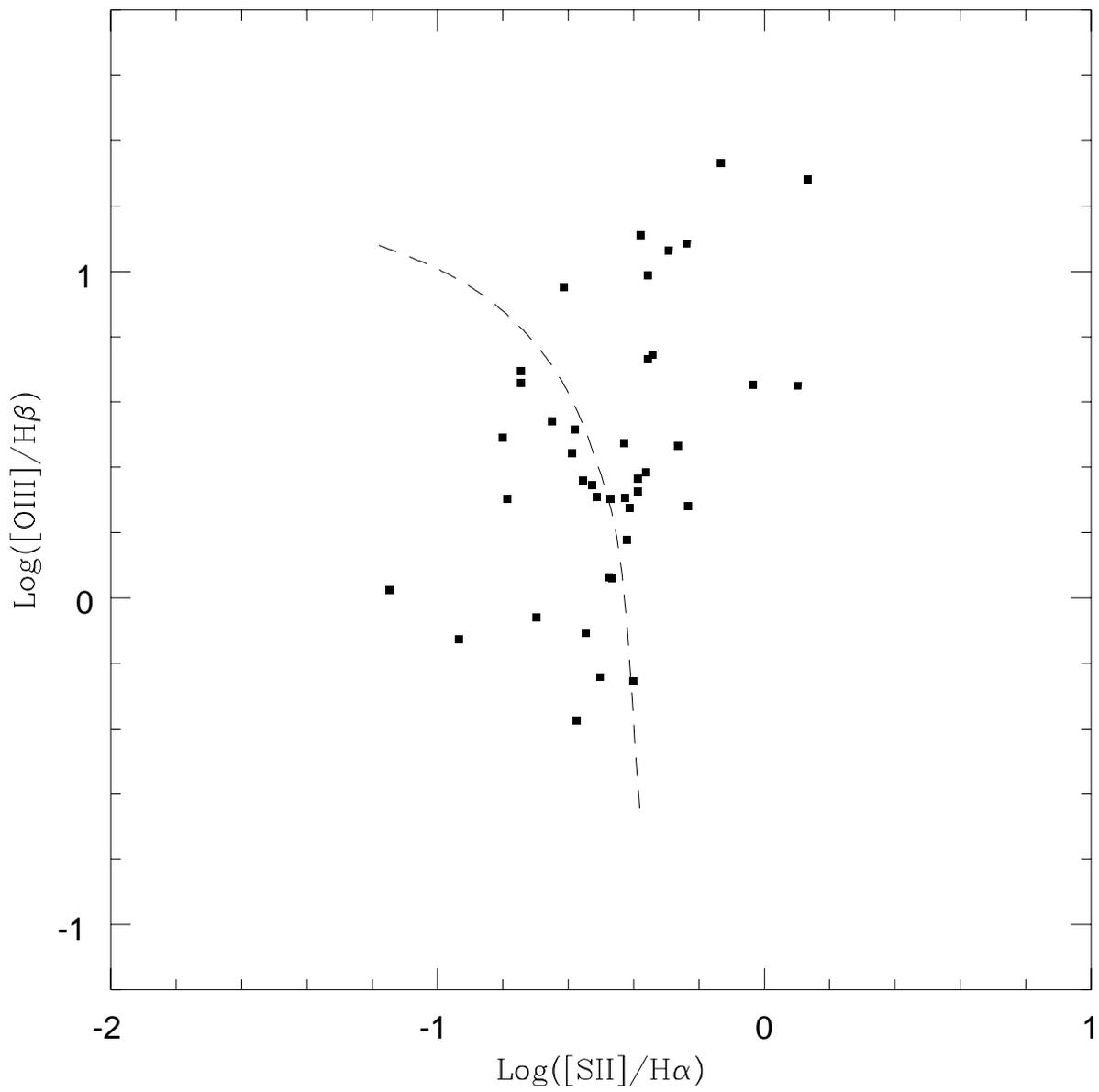
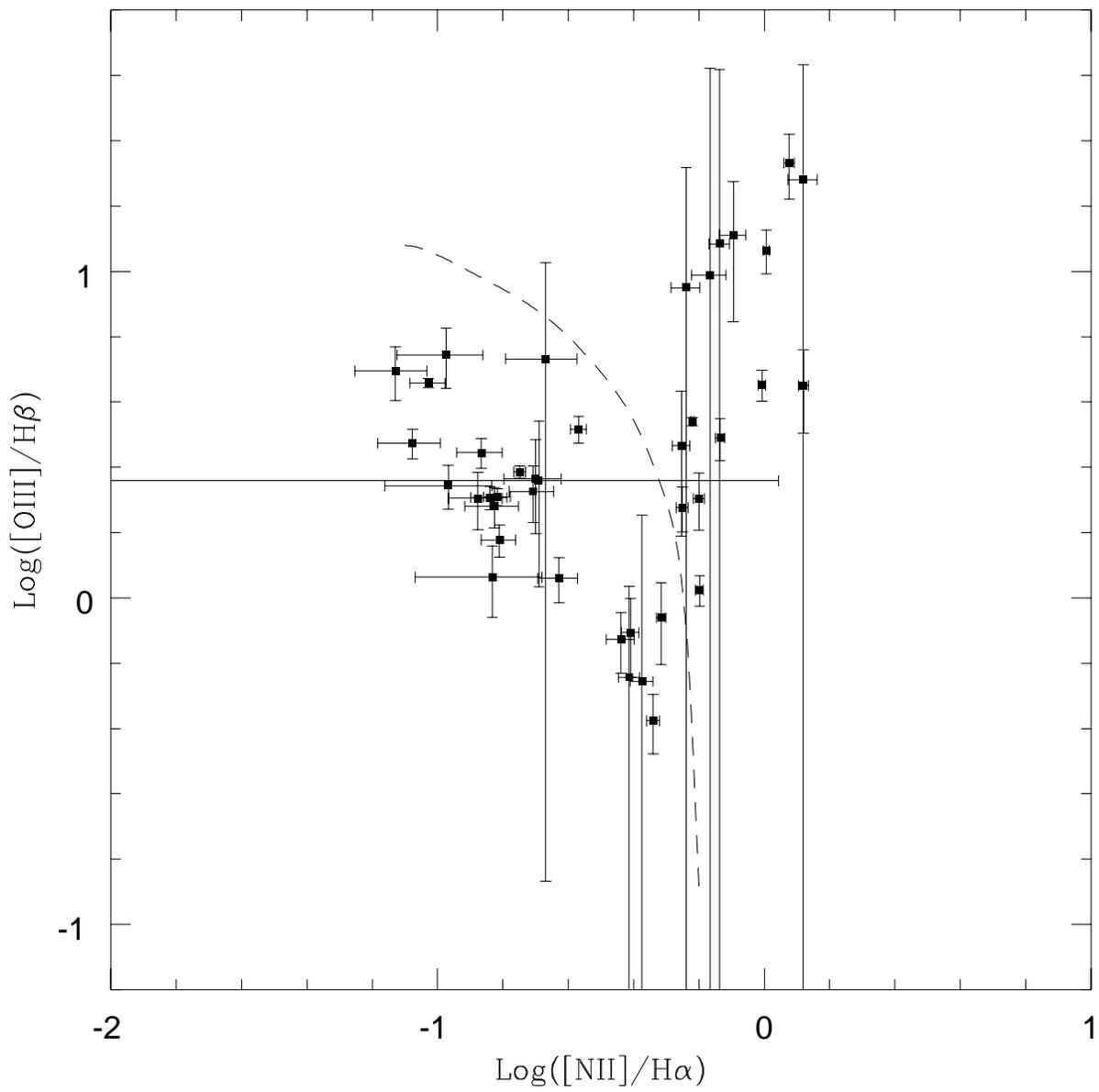
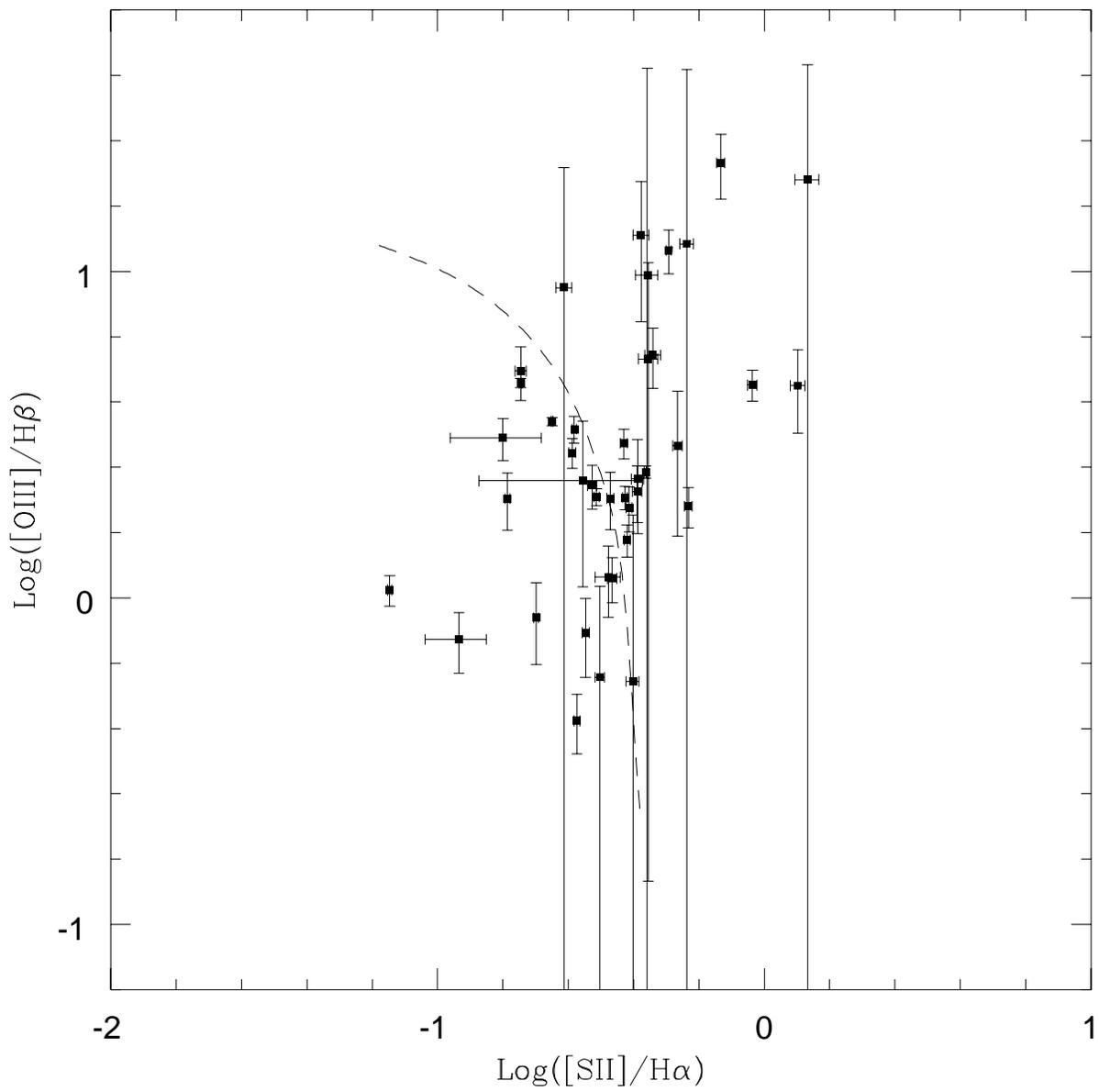
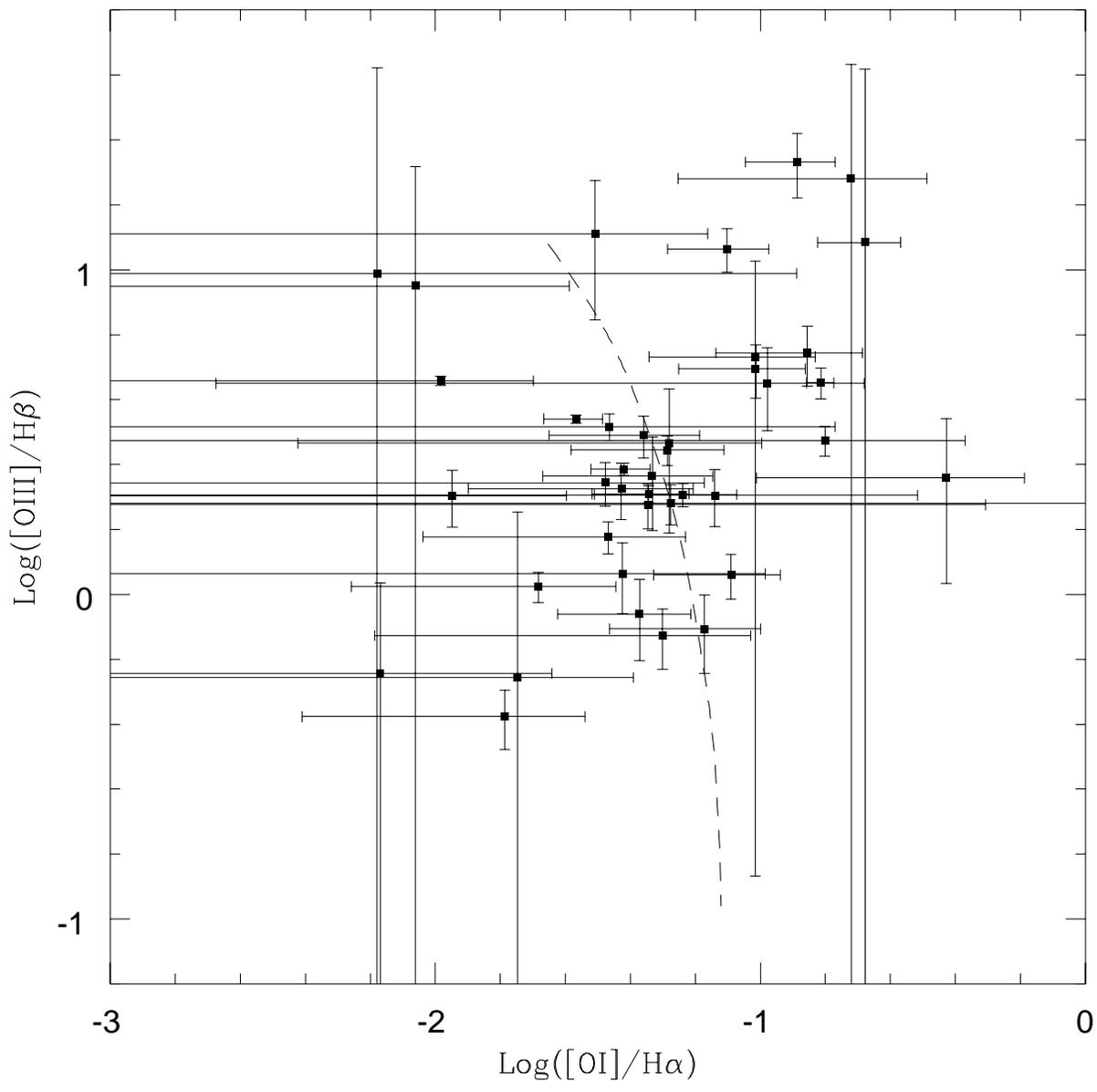
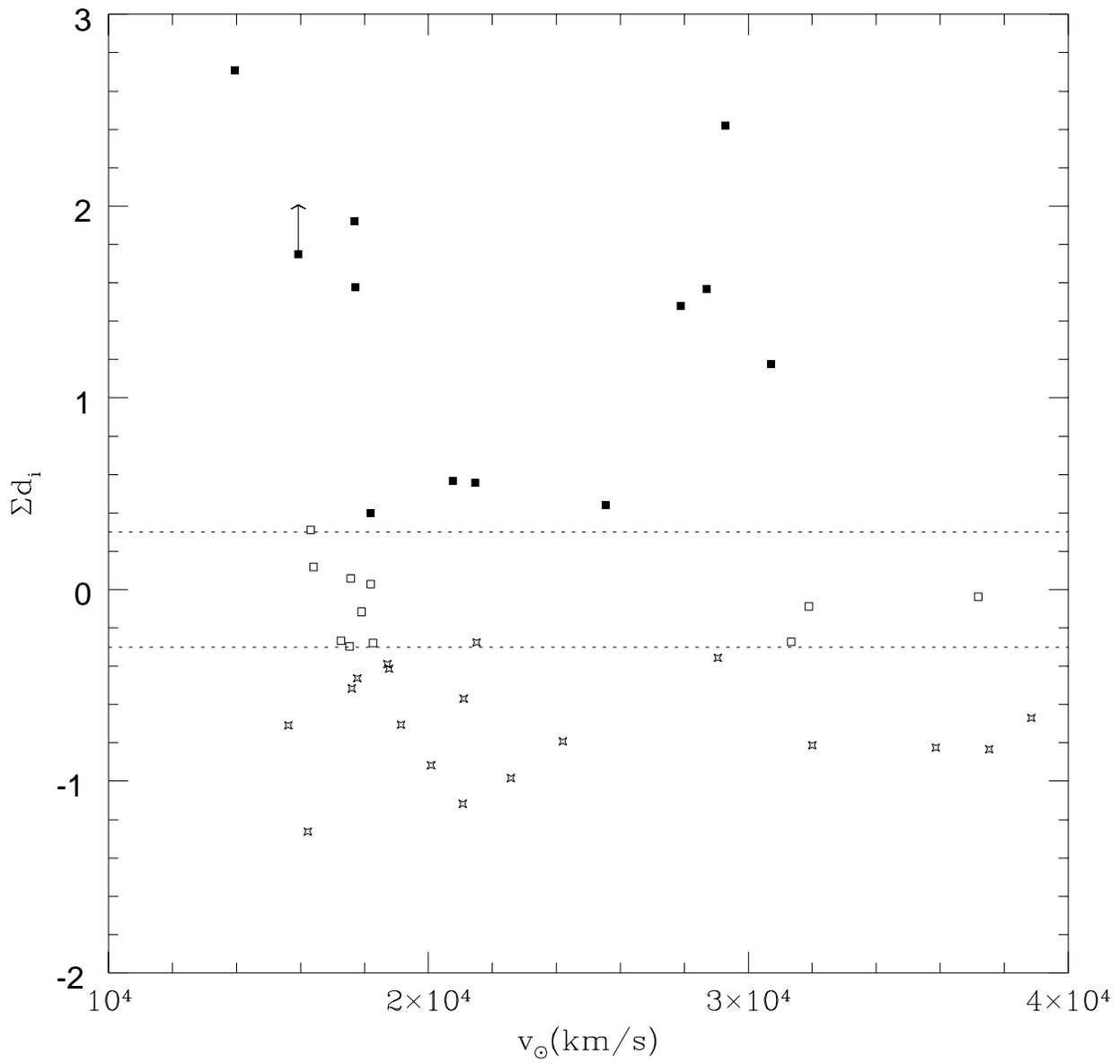
\end{document}